\documentclass{aa}

\usepackage{txfonts}
\usepackage{graphicx}
\usepackage{natbib}
\usepackage{aalongtable}
\usepackage{url}
\bibpunct[; ]{(}{)}{,}{a}{}{;} 
\usepackage{subcaption}
\usepackage{appendix}
\usepackage[normalem]{ulem}
\usepackage{gensymb}
\usepackage{hyperref}
\usepackage{rotating}
\usepackage[normalem]{ulem}
\usepackage{soul}

\begin{document}

\title{LBV phenomenon and binarity: The environment of HR~Car\thanks{Based on observations made with ESO Telescopes at the La Silla Paranal Observatory under programme ID 0102.C-0589(A).}
}
\titlerunning{LBV phenomenon and binarity: the environment of HR~Car} 

\author{
A.~Mehner\inst{1}
\and S.~Janssens\inst{2}
\and C.~Agliozzo\inst{3}
\and W.-J.~de~Wit\inst{1}
\and H.~M.~J.~Boffin\inst{3}
\and D.~Baade\inst{3}
\and J.~Bodensteiner\inst{2}
\and J.~H.~Groh\inst{4}
\and L.~Mahy\inst{5}
\and F.~P.~A.~Vogt\inst{6}
} 

\institute{ESO -- European Organisation for Astronomical Research in the Southern Hemisphere, Alonso de Cordova 3107, Vitacura, Santiago de Chile, Chile 
 \and Institute of Astronomy, Celestijnenlaan 200D bus 2401, 3001 Leuven, Belgium
 \and ESO -- European Organisation for Astronomical Research in the Southern Hemisphere, Karl-Schwarzschild-Strasse 2, 85748 Garching, Germany
 \and Trinity College Dublin, The University of Dublin, Dublin, Ireland
 \and Royal Observatory of Belgium, Avenue Circulaire 3, B-1180 Brussel, Belgium
 \and Federal Office of Meteorology and Climatology MeteoSwiss, Chemin de l'Aérologie 1, 1530 Payerne, Switzerland
}   

\date{Received 2 November 1992 / Accepted 7 January 1993}

\abstract {
Luminous blue variable stars (LBVs) are of great interest in massive-star evolution as they experience very high mass-loss episodes within short periods of time. HR~Car is a famous member of this class in the Galaxy. It has a large circumstellar nebula and has also been confirmed as being in a binary system. 
One means of gaining information about the evolutionary status and physical nature of LBVs is studying their environments. We investigated the stellar content within $\sim100$~pc of HR~Car and also its circumstellar nebula. Very Large Telescope (VLT) Multi Unit Spectroscopic Explorer (MUSE) observations of a $2\arcmin \times 2\arcmin$ region around the star highlight the incompleteness of stellar classification for stars with magnitudes of $V>13$~mag. Eight B0 to B9 stars have been identified which may lie in close spatial vicinity to HR~Car.
For a region with a radius of $r =1.2\degree$ ($\sim 100$~pc at a distance of 4.8~kpc)  around HR~Car, existing catalogs list several late O-type and early B-type stars, but only one early O-type star. Given the relatively low stellar and nebular masses in the HR~Car system, no early O-type stars and only a few late O-type stars would be expected in association with HR~Car. Instead, HR~Car's location in a point vector diagram suggests that HR~Car is not isolated, but is part of a moving group with a population of B-type stars in a spiral arm, and it has not received a strong kick from a supernova explosion of a companion star or a merger event.
Potential binary evolution pathways for the HR~Car system cannot be fully explored because of the unknown nature of the companion star.  Furthermore, the MUSE observations reveal the presence of a fast outflow and ``bullets'' that have been ejected at intervals of about 400 years. These features may have been caused by recurrent mass transfer in the system.
} 


\maketitle

\section{Introduction}
\label{intro}

One of the most debated issues in massive-star research is how mass loss and binarity impact stellar evolution. In this context, understanding luminous blue variable stars (LBVs) and their environments is of great interest as they experience high mass-loss episodes and thus represent a critical phase in massive-star evolution. Several physical causes are discussed for the high mass-loss events of LBVs. Most of them invoke the star's proximity to its Eddington limit and models include an opacity-modified Eddington limit, sub-photospheric gravity-mode instabilities, super-Eddington winds, envelope inflation, fast rotation, and binary interactions and mergers (e.g., \citealt{1994PASP..106.1025H,2020Galax...8...10D} and references therein;   \citealt{2009ApJ...705L..25G,2012A&A...538A..40G,2015A&A...580A..20S,2014ApJ...796..121J,2015MNRAS.447..598S}).

Information on the evolutionary status and physical nature of LBVs can be gained by investigating their environments and circumstellar nebulae with integral field unit (IFU) spectroscopy. IFU observations facilitate studies of the stellar content around LBVs and of the large circumstellar nebulae found around most LBVs. They enable one to identify stellar clusters and to study the massive stellar content and feedback, but also to determine gas and dust properties of the nebulae and to reconstruct past and present-day wind geometries.

One means of investigating the evolutionary status of LBVs is by studying their spatial association with O- and B-type stars. 
In traditional single-star evolutionary models,  high-luminosity LBVs ($log~L/L_\odot > 5.7$, $M_\mathrm{init} > 40~M_\odot$) are brief ($\sim 10^5$~yr) transitions between single massive O-type and Wolf-Rayet (WR) stars  \citep{1984IAUS..105..233C, 1994PASP..106.1025H,2014A&A...564A..30G}.   Less luminous LBVs ($log~L/L_\odot < 5.7$, $M_\mathrm{init} \sim 20-40~M_\odot$) are assumed to be post-red supergiants \citep{1988ApJ...324..279L,1994PASP..106.1025H}.  Recent models show that LBVs can also be at the end stage of massive-star evolution, especially at low metallicity \citep{2019A&A...627A..24G}. 

For a typical velocity dispersion in a cluster of a few km~s$^{-1}$, a star would move less than $\sim$10~pc in 3~Myr, which is the typical lifetime of very massive stars. Thus LBVs and O-type stars should be co-spatial if they evolve as single-star models predict.  
In the past, only a few stellar association studies have been conducted for LBVs. \citet{2003A&A...400..923H} resolved clusters with sizes of about $4~\mathrm{pc} \times 4~\mathrm{pc}$ around RMC~127 and RMC~128 in the Large Magellanic Cloud (LMC). They found that the brightest members around RMC~128 are evolved massive stars with late-O and early-B spectral types. The cluster around the LBV RMC~127 contains fewer members and no stars with spectral types earlier than B0. 

The topic of a single- versus binary-star evolutionary path for LBVs has been developing into a lively scientific discussion.  
\citet{2015MNRAS.447..598S} and \citet{2016MNRAS.461.3353S,2019MNRAS.489.4378S} found that LBVs reside in relatively isolated environments  and are not concentrated in young massive clusters with early O-type stars, as would be expected according to the traditional view of single-star evolution. The authors reason that LBVs are primarily the product of binary evolution. A strong motivation for this hypothesis is the finding that binary interaction dominates the evolution of massive stars (e.g., \citealt{2012Sci...337..444S}). Smith et al.\ propose that most WR stars and Type Ibc supernova (SN) progenitors are mass donors, while LBVs are mass gainers. Through mass transfer, rejuvenated mass gainers get enriched, spun up, and sometimes kicked far from their birth sites by a SN of their companion.
\citet{2017MNRAS.472..591A} derived an approximate analytical model for the passive dissolution of young stellar clusters and computed the relative separation between O-type stars and LBVs. They found that the standard single-star evolution model is mostly inconsistent with observed LBV environments, further suggesting that LBVs are rejuvenated stars in binary systems. One scenario is that they increase their lifetime as the result of mergers, another is that they are mass gainers and receive a kick when the primary star explodes either by an asymmetric explosion or the Blaauw mechanism \citep{1961BAN....15..265B}.

\citet{2016ApJ...825...64H} tested the hypothesis of \citet{2015MNRAS.447..598S} for the LBVs in M31, M33, and the Magellanic Clouds. They show that LBVs are associated with luminous young stars and supergiants that are appropriate for the LBV luminosities and positions in the Hertzsprung–Russell  (HR) diagram. The more luminous LBVs have a spatial distribution similar to late O-type stars, while the less luminous LBVs that are assumed to be post-red supergiants have a distribution similar to red supergiants (see also \citealt{2016arXiv160802007D}).  The spatial velocities of LBVs are also consistent with their positions in the respective galaxies.
\citet{2018AJ....156..294A} emphasize that our knowledge of the spectroscopic content of the LMC, M31, and M33 is quite incomplete and instead used photometric criteria to select the highest-mass unevolved stars from spatially complete photometric catalogs of these galaxies. They found that the majority of LBVs are in or near OB associations as are the blue supergiants and WR stars, while the red supergiants are not. They conclude that the spatial distribution of LBVs is consistent with a single-star evolutionary path. \citet{2019MNRAS.489.4378S} argue that bright blue stars are not good tracers of the youngest massive O-type stars and uphold their  proposal of LBVs being binary products. 

Recently, \cite{2021arXiv210512380M} provided a bias-corrected spectroscopic binary fraction among Galactic (candidate) LBVs of $62^{+38}_{-24}$\%, covering period ranges of $1-1000$~d. They found an even higher binary fraction of 78\% using interferometry for projected separations up to $100-150$~mas. Stellar radii of $100-650~R_{\odot}$ were also derived in this work. Taken together, these characteristics make short-period binary systems among (candidate) LBVs unlikely. If LBVs form through single-star evolution, the initial orbits must be wide. On the other hand, if they are the end products of binary evolutionary channels, then massive stars in a short-period binary system either undergo a phase of fully nonconservative mass transfer to widen the orbit or LBVs form through mergers in binary or triple systems. 

We aim to set the stage for further studies of the population of LBVs by investigating the stellar content of the volume around the Galactic LBV HR~Car and its circumstellar nebula. HR~Car's nebula has a bipolar morphology with two lobes of $\sim19\arcsec$ in diameter \citep{1997ApJ...486..338N}. Nebular mass estimations range from $0.8~M_{\odot}$ \citep{2000ApJ...539..851W} to $2.1~M_{\odot}$ \citep{1995AJ....110..251C} for the ionized gas, which can be seen as a proxy for the total nebular mass.
The star had two recent outburst events, in 1989--1993 and in 1997--2005, when it moved temporarily across the HR diagram to cooler temperatures \citep{ 2003IAUS..212..243S}.\footnote{See also the American Association of Variable Star Observers (AAVSO) lightcurve at \url{https://www.aavso.org}.} 
The hypothesis that LBVs are mass gainers, and at least originally the less massive component of massive binaries, can be tested in particular with HR~Car as it is both an LBV and a binary.

Interferometric monitoring of HR~Car, complemented by HARPS spectra show an orbital period of $P=2330$~d, an eccentricity of $e=0.106$, a minimum separation between the stars of about 10~au, an inclination of $i=119\degree$, a mass of the primary of $\sim 21~M_{\odot}$, and a companion mass of $\sim 10~M_{\odot}$, assuming a distance of $d=4.9$~kpc  (Boffin et al., in prep.).\footnote{These preliminary orbital parameters are much improved compared to the best-fit orbit presented in \citealt{2016A&A...593A..90B}.} The nature of the companion is still unknown.  \citet{2016A&A...593A..90B} suggested that the secondary could be a red supergiant, but no corresponding spectral features were observed in a near-infrared spectrum obtained in 2017 (Boffin et al., in prep.). The current stellar and nebular masses imply that the LBV in HR~Car, and probably the initial main-sequence masses of the binary or triple  components, barely exceeded the mass of an O-type star. Given the evolved nature of HR~Car, one would not expect (higher-mass) O-type stars to be still around in its vicinity. We thus investigate if HR~Car is part of a moving group, that is part of a stream of stars with common age and motion.

In Section  \ref{data} we describe the observations, data analysis, and auxiliary data. Section \ref{environment_HRCar} presents the stellar environment of HR~Car. Section \ref{circumstellarnebula} discusses the large-scale circumstellar nebula and Section \ref{bipolaroutflow} the discovery of a fast inner outflow and ``bullets''. In Section \ref{environment_LBVs} we place these results in a larger framework of LBV evolution. We conclude with Section \ref{conclusion}.

\section{Observations and auxiliary data}
\label{data}

\begin{figure}
\centering
\resizebox{1\hsize}{!}
{\includegraphics[width=1\textwidth]{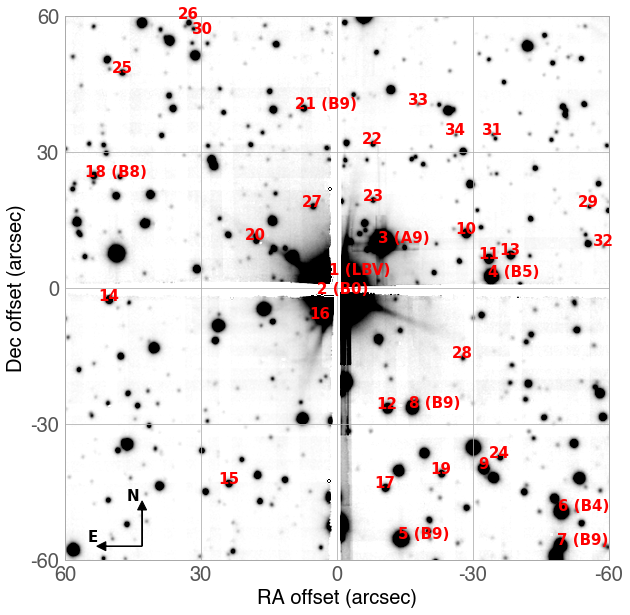}}
     \caption{Collapsed $V$-band image from the MUSE datacubes of the four deep pointings separated by small gaps, mapping the region $2\arcmin \times 2\arcmin$ ($2.8~\mathrm{pc} \times 2.8~\mathrm{pc}$ at a distance of 4.8~kpc) around HR~Car in the center, labeled ``1''. The object numbering is according to Table \ref{table:spectraltypes}. Approximate limiting magnitude is $V\sim22$~mag.}
     \label{figure:MUSEdatacube}
\end{figure}

\begin{figure*}
\resizebox{1\hsize}{!}
{\includegraphics[width=1\textwidth]{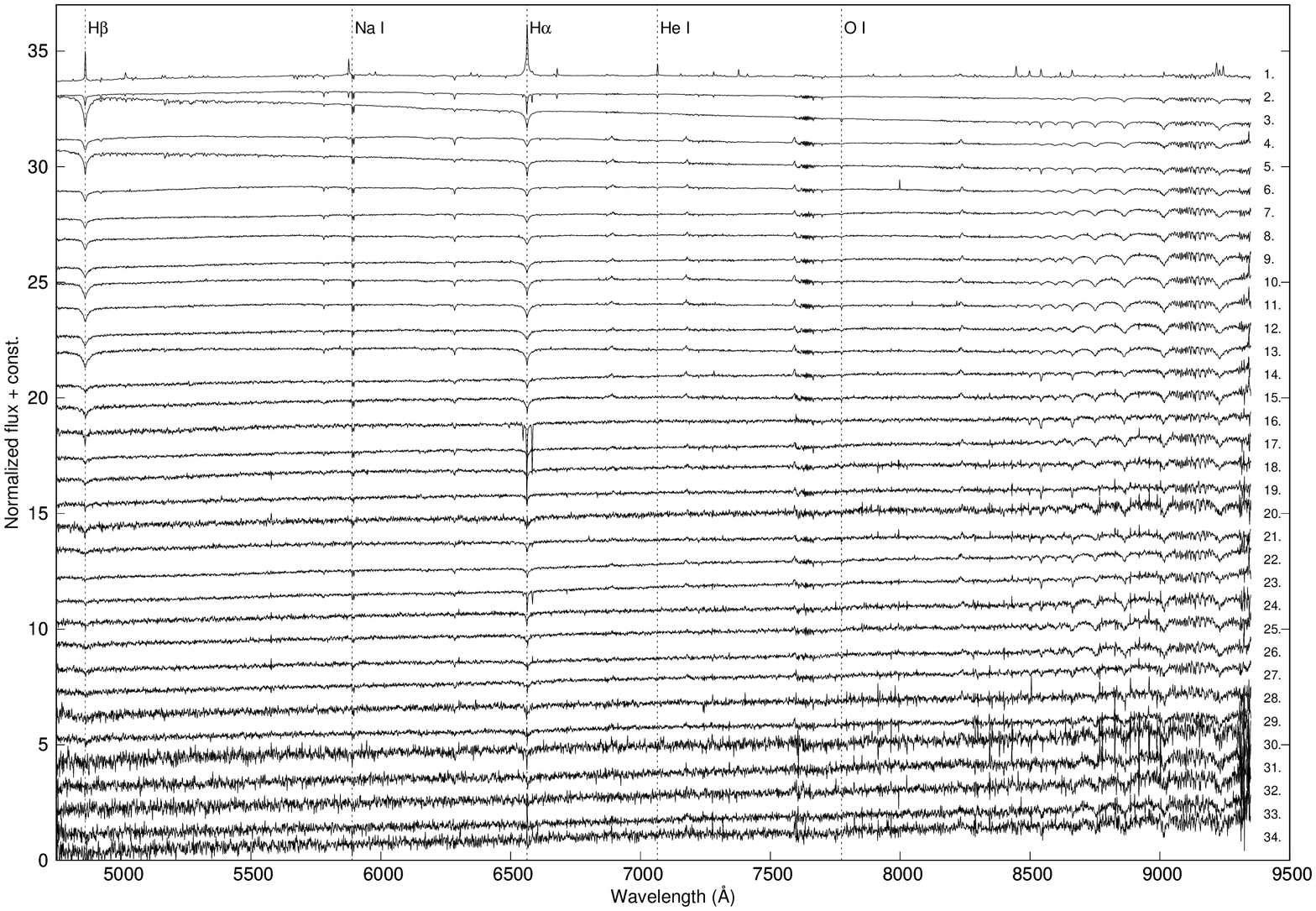}  } 
\caption{Flux-calibrated spectra of hot stars in the $2\arcmin \times 2\arcmin$ region around HR~Car in the MUSE data, see Table~\ref{table:spectraltypes} for details. The sources are ordered by increasing $V$-band magnitude. The spectral energy distribution of the stars 3., 5., and 13.\ indicate that they are foreground stars. The low spectral resolving power of the MUSE spectra prevents detailed radial velocities measurements of these sources.
}
     \label{figure:MUSEspectra}
\end{figure*}

\subsection{IFU observations}
Observations of HR~Car with the Multi-Unit Spectroscopic Explorer (MUSE;  \citealt{2010SPIE.7735E..08B}) at UT4 of ESO's Very Large Telescope (VLT) at Paranal  in its seeing-limited Wide Field Mode (WFM) were obtained on 28, 30, and 31 December 2018.
MUSE is an integral field spectrograph, composed of 24 identical IFU modules that together sample a $1\arcmin \times 1\arcmin$  field of view in WFM. Spectrally, the instrument covers the wavelength range $480-930$~nm with a resolving power of $R \sim 2000-3500$, which allows to perform spectral classification and synthetic photometry, as, for example, demonstrated in \citet{2016A&A...585A..81M}. The WFM has a spatial sampling of 0.2\arcsec\ per pixel.

We mapped the environment of HR~Car with 5 pointings. One pointing, with HR~Car at the center, was observed with 48 exposures of 15s to obtain a deep image of HR~Car's circumstellar nebula, while avoiding saturation and strong straylight artifacts by the bright central source. Four additional pointings map a region of $2\arcmin \times 2\arcmin$ around HR~Car with two exposures of 300s per pointing (Figure \ref{figure:MUSEdatacube}). The obtained image quality at 600~nm is about 1\arcsec. The data were reduced with the ESO MUSE pipeline version 2.8.3 \citep{2020A&A...641A..28W,2016ascl.soft10004W}, using the option to obtain collapsed images in several  filters for photometry (e.g., Johnson $V$, Cousins $I$, and SDSS $i$).

\subsection{Auxiliary data}

Using a simple inversion of the parallax from the Gaia Early Data Release 3 (Gaia EDR3\footnote{\url{https://gea.esac.esa.int/archive/}}; \citealt{2018A&A...616A...1G,2016A&A...595A...1G}), the distance to HR~Car is $d=4922\pm403$~pc. This method is normally warranted, given the small error bar. Using a Galaxy model prior, \citet{2021AJ....161..147B} estimated a geometric distance to HR~Car of $d=4752$~pc, with an uncertainty range of $4448-5098$~pc. These distance estimates are slightly lower than previously published distances such as a kinematic distance of $5.4\pm0.4$~kpc \citep{1991A&A...248..141H} or a distance of $5\pm1$~kpc derived via multicolor photometry of field stars around HR~Car  \citep{1991A&A...246..407V}, but within the uncertainties. The distance determination from the parallax may be compromised by the binarity and bright circumstellar nebula.  \citet{2019MNRAS.488.1760S} used Bayesian inference to estimate the distance from Gaia DR2 data and determined a distance of $d=4.37$~kpc ($3.50-5.72$~kpc).

For the analysis of the stellar content around HR~Car, we queried Simbad \citep{2000A&AS..143....9W} for stars in a region with radius $1.2\degree$ around HR~Car, which corresponds to an area with radius $100$~pc at a distance of 4.8~kpc, typical for an OB association \citep{2020NewAR..9001549W}. No overdensity of stars is discernible in this area on sky as would be expected in case of a stellar association. The HR~Car binary is also too low in mass for its evolved state to still expect many O-stars in its neighborhood. We thus analyze this dataset to potentially identify a moving group around HR~Car, that is stars with common age and motion.
Throughout the paper, we use Gaia EDR3 proper motions and parallaxes and the geometric distance estimates from \citet{2021AJ....161..147B}. 

\subsection{Photometric calibrations}
\label{dataanalysis}

We used the tool QFitsView \citep{2012ascl.soft10019O} to analyze the stellar content and nebula around HR~Car. From the MUSE data cubes, we extracted spectra of all objects with a sufficient signal-to-noise ratio (S/N $> 50$) to determine their spectral types. 
We derived photometry on the collapsed images using SExtractor \citep{1996A&AS..117..393B}. For the Johnson $V$, Cousins $I$, and SDSS $i$ filters, 100\% of the spectral region is covered by MUSE. 

The MUSE observations were obtained in cloudy sky conditions with transparency variations above 20\% and none of the sources in the observed fields (apart from HR~Car) have published magnitudes in Simbad.\footnote{One of the objects is listed in Simbad (GSC~08612-01828; spectral type A9). However, the reference \citep{1991A&A...246..407V} reveals that the star has been misidentified with a different star outside the field observed with MUSE.} We cross-match the sources detected in the MUSE images collapsed over the SDSS $i$ passband with the SkyMapper Southern Survey \citep{2019PASA...36...33O} and apply the magnitude offsets found for each pointing. The errors of these magnitude offsets are between $0.1-0.3$~mag, depending on the pointing. Comparison of a few stars that were observed in different pointings (on different observing dates and with different exposure times) shows that the derived photometry is better than 0.2~mag across pointings and dates.

In the resulting MUSE photometry (Table~\ref{table:spectraltypes}), HR~Car is  with $V=8.85$~mag about $0.1$~mag fainter in $V$ and  with $I=7.54$~mag about $0.3$~mag fainter in $I$ band compared to the AAVSO lightcurve, within the error of the cross-match with the SkyMapper Southern Survey. Additional uncertainties arise from the different image qualities between the data sets, different aperture photometry methods and aperture sizes employed, and potentially slightly different filter transmission curves.

\section{The environment of HR~Car}
\label{environment_HRCar}

\begin{figure*}
\centering
\resizebox{1\hsize}{!}
{\includegraphics[width=1\textwidth]{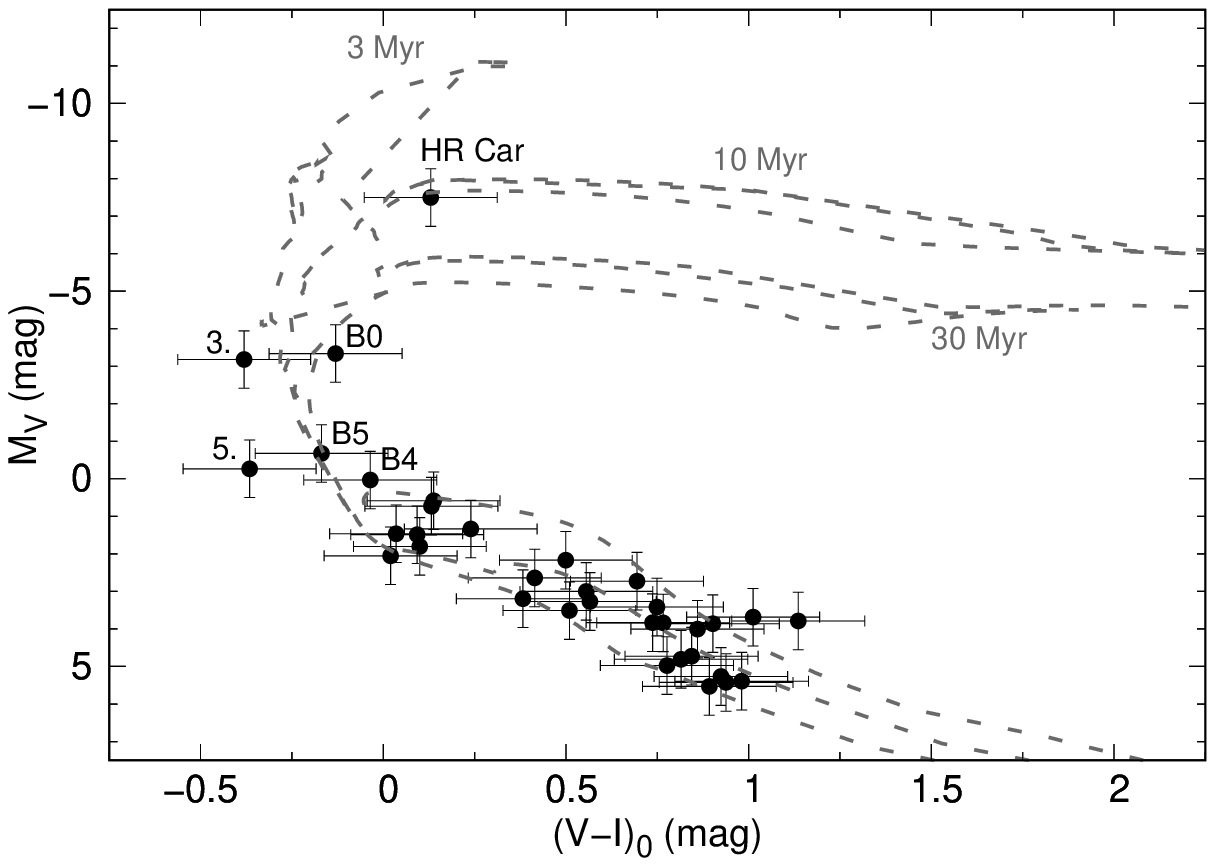}
\includegraphics[width=1\textwidth]{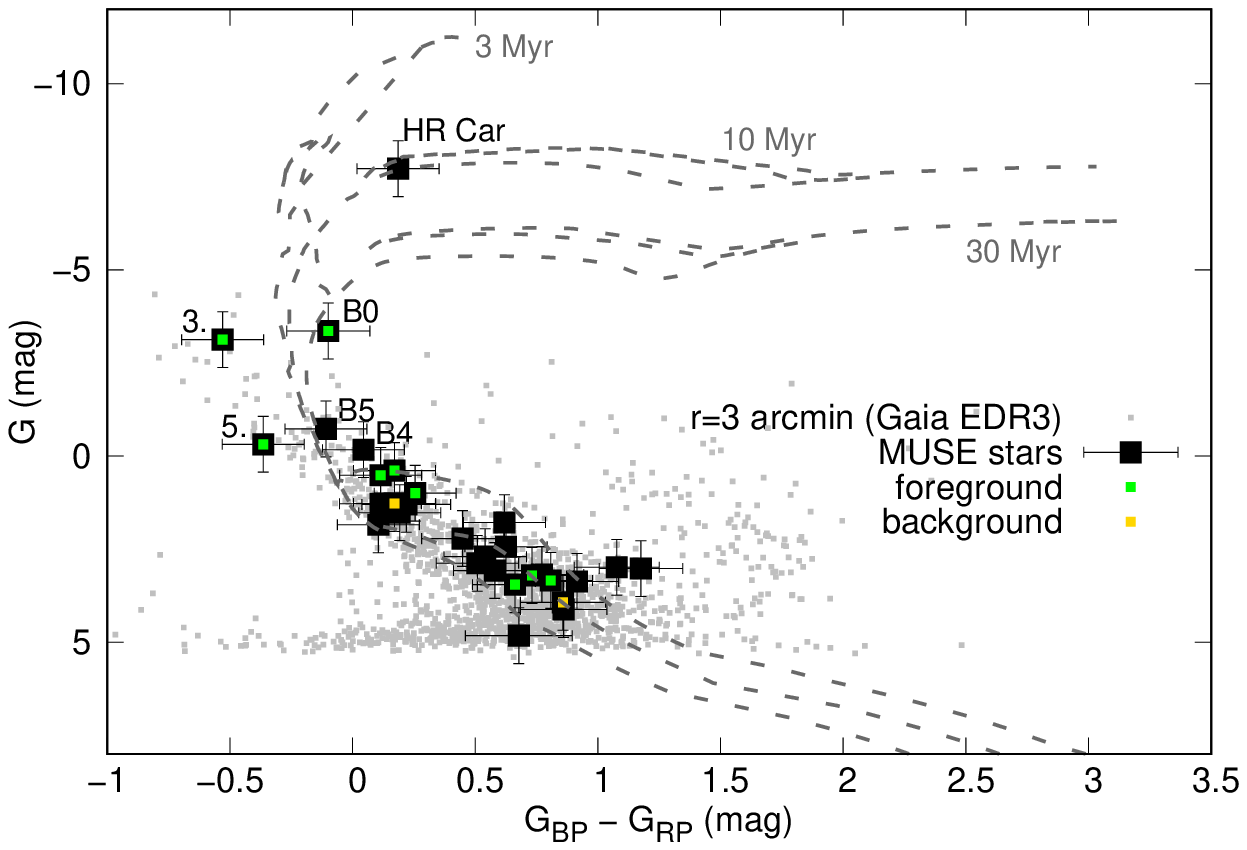}}
     \caption{{\it Left:} Color-magnitude diagram of the hot stars within a $2\arcmin \times 2\arcmin$ region around HR~Car derived from MUSE data and synthetic photometry, assuming a distance of $d=4.8$~kpc to all sources and corrected for an extinction of $A_V=2.9$~mag. The errors are dominated by the uncertainties of the reddening and distance estimates. PARSEC theoretical single-star stellar evolution isochrones are shown for 3~Myr, 10~Myr, and 30~Myr (dashed gray curves; \citealt{2012MNRAS.427..127B}). The stars 3.\ and 5.\ are foreground stars.  {\it Right:}  Same as left figure, but with Gaia EDR3 photometry. Stars with estimated higher geometric distance limits of $d_\mathrm{geo\_h}  < 4448$~pc are indicated as foreground stars (green squares) and stars with estimated lower geometric distance limits of $d_\mathrm{geo\_l} > 5098$~pc as background stars (golden squares). The other sources (black filled squares only) are at an inferred distance compatible with that of HR~Car. The gray small squares show all Gaia sources within an area of radius $r=3\arcmin$ from HR~Car.
}
     \label{figure:HRdiagram}
\end{figure*}

\begin{figure*}
\centering
\resizebox{1\hsize}{!}
{\includegraphics[width=0.32\textwidth]{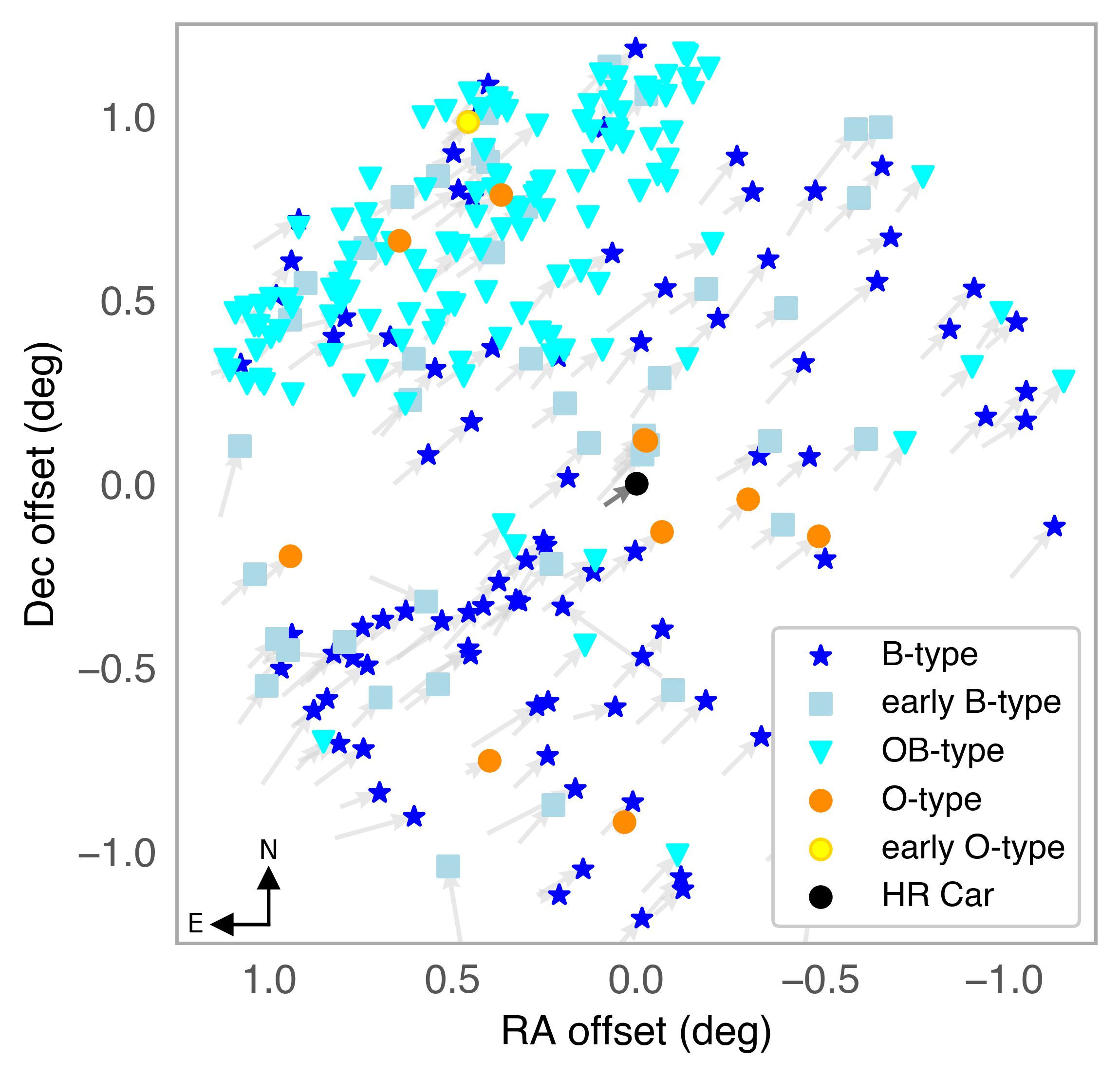} 
\includegraphics[width=0.35\textwidth]{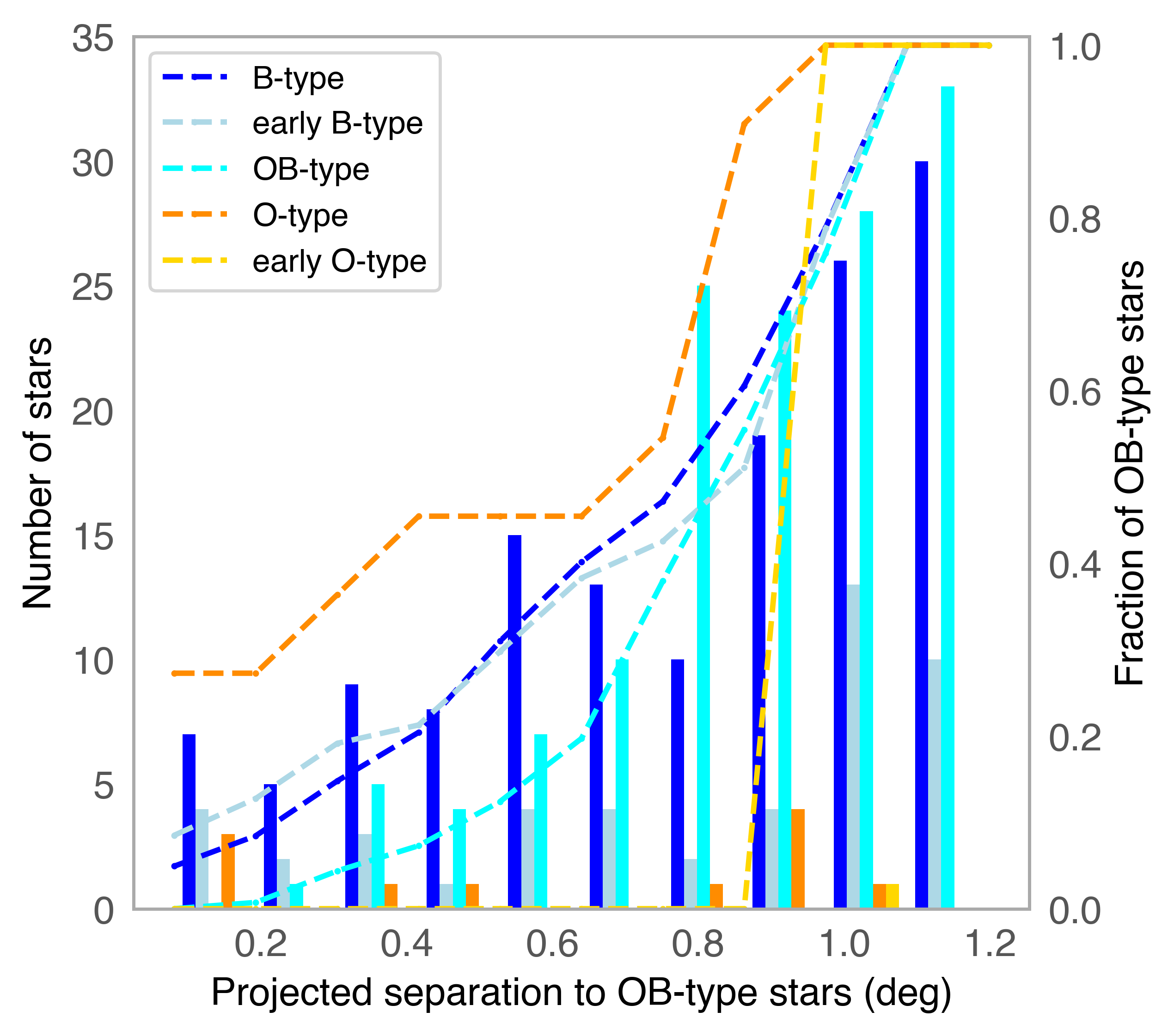}
\includegraphics[width=0.305\textwidth]{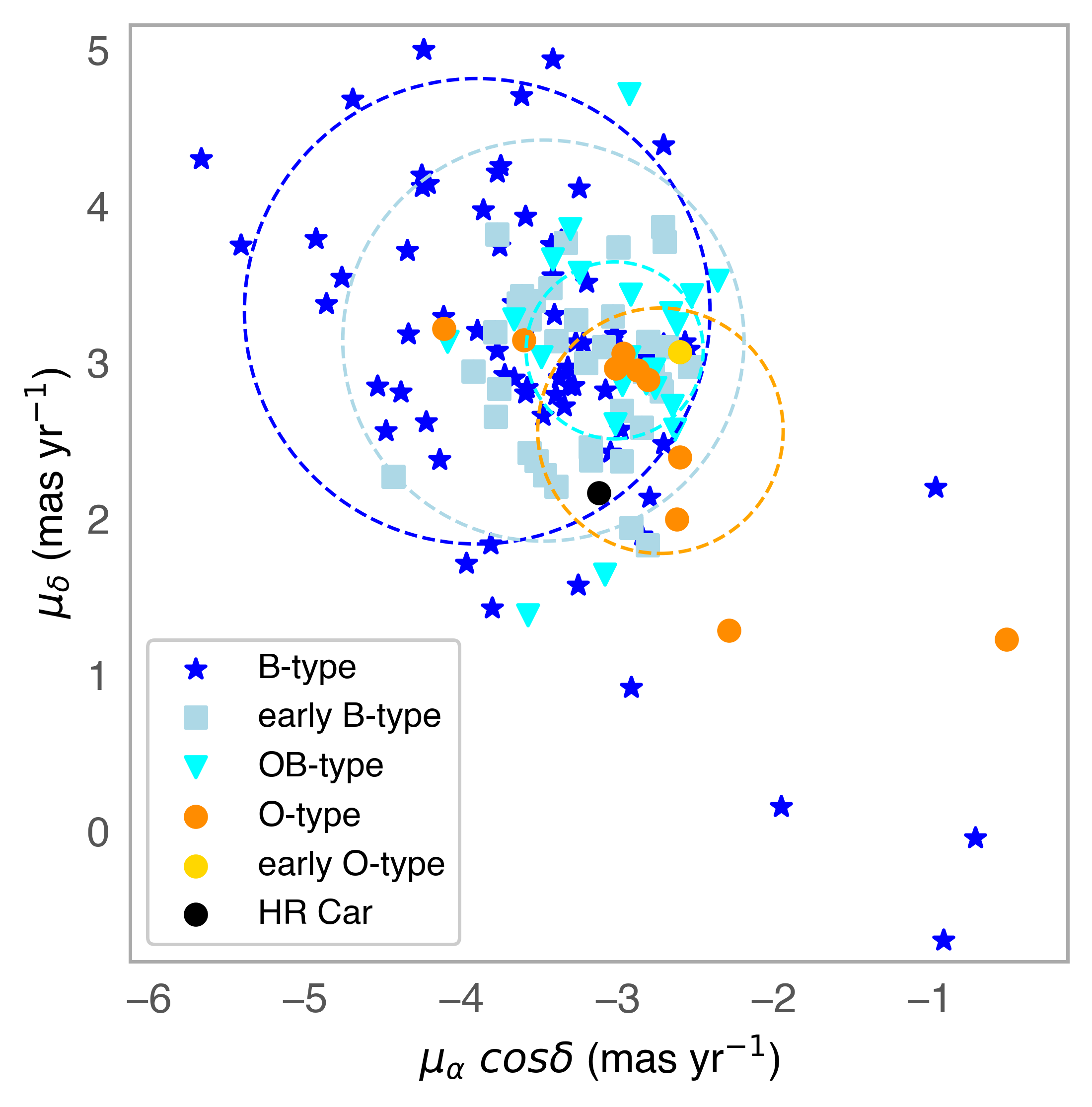} } 
\caption{{\it Left:} Spatial distribution of all O- and B-type stars in a projected region with radius $r=1.2\degree$ around HR~Car (black filled circle in the center), based on information retrieved from Simbad and Gaia. The newly classified B-type stars  in the MUSE data are not listed in Simbad and not included of this figure. The arrows indicate the movement of the stars in the past $10^5$~yr, inferred from available Gaia proper motions. HR~Car moves in the same general direction as other massive stars in this region of the sky. {\it Middle:} Projected separation to O- and B-type stars and the normalized cumulative distribution (dashed curves). {\it Right:} Vector point diagram for O- and B-type stars using Gaia proper motions. The dashed circles indicate two standard deviations of the mean of the proper motions for the respective stellar subgroups. HR~Car's location (black filled  circle at the center of the plot) is within two standard deviations of the mean of the proper motions for O-type, early B-type, and B-type stars.}
     \label{figure:HRclusterSimbad}
\end{figure*}

\subsection{Spectral typing and extinction determination}
\label{spectraltyping}

To investigate the hot stellar content in the close environment of HR~Car, we visually inspected the MUSE spectra and selected the objects with broad Balmer and Paschen hydrogen lines and HeI~7065 for the earliest spectral types. This resulted in 34 objects, see Figure~\ref{figure:MUSEspectra} and Table~\ref{table:spectraltypes}.
Stellar spectral types were determined by comparing the equivalent widths of the HeI~$\lambda$7065, H$\beta$, NaI~$\lambda$5889 (stellar component), and OI~$\lambda$7774 lines with those tabulated as function of spectral type in \citet{1995bces.book.....J}. In addition, we used the stellar spectral classification software PyHammer 2.0.0 to compare the MUSE spectra to optical spectral templates and spectral line index measurements \citep{2020ApJS..249...34R,2020ascl.soft02011K}. 
We newly classified eight stars as B-type within the $2\arcmin \times 2\arcmin$ region around HR~Car. Three of these objects are early B-type stars with spectral classifications of B0, B4, and B5, respectively. There is no O-type star in this region of the sky. The spectral resolution and S/N of the spectra is not conducive for detailed stellar atmosphere studies, such as $log~g$ determination.  

The AAVSO lightcurve shows that HR~Car has been in a quiescent, hot state since about 2010. This is also confirmed by optical spectra obtained in the past decade with multiple instruments, for example, our 2018 MUSE spectrum. From the MUSE photometry of HR~Car, $V=8.85$~mag, an absolute magnitude of $M_\mathrm{V} = -7.5$~mag is derived, assuming $d=4.8$~kpc, $E(B-V) = 0.95$~mag (see below), and $R_V = 3.1$. The best-fit CMFGEN model of a 2017 VLT/X-shooter spectrum suggest an effective temperature of $T_\mathrm{eff}=17\,750$~K (Groh et al., in prep.). With a bolometric correction of -1.6~mag \citep{1996ApJ...469..355F}, the bolometric magnitude of HR~Car is $M_{bol}= -9.1$~mag. During HR~Car's outburst, that is cool state, its temperature has been estimated to about $8\,000$~K \citep{2001A&A...366..508V}. A magnitude of $V \sim 7$~mag (AAVSO lightcurve) and a bolometric correction of about 0~mag, results in a bolometric luminosity comparable to its quiescent state. Groh et al. (in prep.) will present a detailed spectroscopic analysis of HR~Car's hot and cool states.  
Adjusted for the different distances used, our luminosity estimate for HR~Car is comparable to previous literature values \citep{ 1988ApJ...324.1071M,1991A&A...246..407V,1991A&A...248..141H,2001A&A...366..508V,2007ApJ...659.1563G,2009ApJ...705L..25G,2010AN....331..349H}.

We employed different methods to determine the extinction toward the 34 MUSE sources: 1) equivalent width of the diffuse interstellar band at $\lambda$5780~\AA,  as described in \citet{2016A&A...585A..81M}, and 2) comparing the relative flux-calibrated MUSE spectra to template spectra. 
The extinction and reddening toward an object depend on the wavelength, the amount of dust along the line of sight, and the dust grain sizes and composition. For HR~Car, using a 2017 X-shooter spectrum and CMFGEN modeling, Groh et al.\ (in prep.) determined a ratio between total and selective extinction of $R_V = 4.3\pm0.1$, which implies large grain sizes in its circumstellar environment. For consistency in our analysis, we adopt $R_V = A_V /E(B-V) = 3.1$ \citep{1989ApJ...345..245C,1999PASP..111...63F} for the interstellar medium toward all stars in the observed MUSE field.

For most of the brighter MUSE sources, the derived reddening is in the range of $E(B-V) = 0.85-1.10$~mag (Table \ref{table:spectraltypes}). This hints at similar distances as the reddening is mostly due to the dust in the interstellar medium along the line of sight, but see Section \ref{massivestars_HRCar} for a discussion on the distances inferred from the Gaia EDR3 release data. The reddening to HR~Car, estimated to $E(B-V) = 1.24$~mag, may be larger due to a significant amount of circumstellar dust. For the faintest stars the derived reddening has a large uncertainty as the spectra are very noisy at the bluest wavelengths, but the larger values may indicate that they lie in the background. 
 For consistency in our analysis, and because of the large uncertainties in the reddening estimations, we adopt a value for the interstellar reddening to HR~Car of $E(B-V) = 0.95\pm0.1$~mag, consistent with $E(B-V) = 0.90$~mag found by  \citet{1991A&A...246..407V} via multicolor photometry of about 100 field stars within a radius of $10\arcmin$ around HR~Car.

\begin{sidewaystable*}
\caption{Hot stellar content in the neighborhood of HR~Car.$^a$ \label{table:spectraltypes}}
\begin{tabular}{llccccccccccccc}
\hline\hline
\# & Coordinates & SDSS$i$ & $I$ & $V$ & E(B-V) & SP & geo & geo\_l & geo\_h & pgeo & pgeo\_l & pgeo\_h & 1/p$^b$ & Gaia source\_id \\ 
&   & (mag) & (mag) & (mag) & (mag) &  & (pc) &  (pc)  &  (pc) & (pc) &  (pc)  &  (pc)  & (pc) & \\ 
        \hline
1. & 155.7244783 -59.6244613 & 7.78  & 7.54 & 8.85 & 1.24 & LBV$^c$ & 4752 & 4448 & 5098 & 4602 & 4269 & 4986   & 4922 &  5255045082580350080   \\
2. & 155.7265119 -59.6245356 & 12.13 & 11.95 & 13.01  & 1.00 & B0 & 4136 & 3869 & 4382 & 4132 & 3927 & 4368 & 4278 & 5255045082561662592\\
3. & 155.7190919 -59.6213844 & 12.56 & 12.36 & 13.17 & 0.20 & A9$^{d,e}$  & 1848 & 1803 & 1895 & 1849 & 1811 & 1884 & 1923 & 5255045151299827456 \\
4. &  155.7057998 -59.6234632 & 14.85 & 14.65 & 15.67 & 0.87 & B5 & 4187 & 3829 & 4539 & 4131 & 3826 & 4549 & 4927 & 5255045151299811584 \\
5. &  155.7167475 -59.6395444 & 15.39 & 15.26 & 16.08 & 0.65 & B9$^{e}$ & 2494 & 2345 & 2728 & 2450 & 2329 & 2630 & 2700 & 5255045048220579456 \\
6. &  155.6973330 -59.6378605 & 15.41 & 15.23 & 16.38 & 1.06 & B4 & 4775 & 4186 & 5492 & 4710 & 4155 & 5401 & 5581 & 5255045013860826880  \\
7. &  155.6974295 -59.6399275 & 15.84 & 15.61 & 16.93 & 1.10 & B9 & 3850 & 3444 & 4393 & 3750 & 3272 & 4242 & 4292 &  5255045013860925824 \\
8. & 155.7153745 -59.6315202 & 16.01 & 15.76 & 17.08 & 1.03 & B9 & 3593 & 3146 & 4202 & 3558 & 3019 & 4277 & 3818 & 5255045116940068352\\
9. &  155.7067251 -59.6352335 & 16.50 & 16.26 & 17.68 & 1.03 & A3 & 3410 & 2892 & 3962 & 3432 & 2987 & 3978 & 3621 &  5255045116940054400 \\
10. &  155.7088481 -59.6208449 & 16.78 & 16.59 & 17.81 & 0.84 & A3 & 5465 & 4007 & 7734 & 4343 & 3548 & 5310 & &  5255045146977490688  \\
11. &  155.7061104 -59.6224064 & 16.75 & 16.56 & 17.84 & 0.84 & A3 & 7045 & 5255 & 9053 & 6391 & 4441 & 8891 &  &  5255045151299814144 \\
12. &  155.7183657 -59.6315608 & 17.07 & 16.86 & 18.15 & 0.95 & A3 & 6570 & 5076 & 9304 & 6447 & 4960 & 8765 &  &  5255045048220593280 \\
13. &  155.7034540 -59.6221636 & 17.40 & 17.19 & 18.40 & 0.84 & A3$^{e}$ & 3901 & 2997 & 5130 & 3537 & 2799 & 4947 & &  5255045146979950336 \\
14. &  155.7521425 -59.6249287 & 17.14 & 16.83 & 18.52 & 0.90 & F3 & 4698 & 3547 & 6373 & 3759 & 3045 & 4576 &  &  5255420393971899136 \\
15. &  155.7376107 -59.6361695 & 17.66 & 17.39 & 18.99 & 1.10 & A1 & 6318 & 4092 & 10492 & 4350 & 3299 & 6143 &  &  5255045078258086400 \\
16. & 155.7265192 -59.6260821 & 17.50 & 17.19  & 19.07 & 0.85 & F4 & 4784 & 3040 & 7220 & 2919 & 2390 & 3630 & &  5255045082561952640 \\
17. &  155.7186046 -59.6364351 & 17.89 &17.60 & 19.35 & 1.42 & A1 & 6997 & 4804 & 10537 & 5632 & 3837 & 8693 & &  5255045043900737152  \\
18. &  155.7539724 -59.6173297 & 18.27 & 17.97& 19.54 & 1.45 & B8 & 4956 & 3312 & 7932 & 4679 & 2997 & 6364 & &  5255420393999646336 \\
19. &  155.7118295 -59.6355545 & 18.14 &17.86 & 19.62 & 1.00 & F2 & 5976 & 3706 & 8379 & 4542 & 3405 & 6132 &  &  5255045112618092032\\
20. &  155.7343382 -59.6211879 & 18.13 & 17.83 & 19.76 & 1.26 & A1 & 977 & 739 & 1370 & 1923 & 1712 & 2236 & &  5255045082577422336 \\
21. &  155.7285538 -59.6131811 & 18.45 & 18.16 & 19.86 & 1.35 & B9 & 4364 & 2685 & 6697 & 3105 & 2281 & 4290 & &  5255420462691375488 \\
22. & 155.7201480 -59.6153487 & 18.21 & 17.83 & 20.03 & 1.35 & F1 & 6884 & 4500 & 9667 & 5724 & 3994 & 9848 &  &  5255045146979952384\\
23. & 155.7201048 -59.6187980 & 18.21 & 17.81 &  20.13 & 1.35 & F1 & 3466 & 2382 & 5316 & 3975 & 2814 & 5570 & &  5255045151281149952 \\
24. & 155.7047429 -59.6345651 & 18.58 &18.26 & 20.18 & 1.00 & F9 & 6272 & 4228 & 8875 & 6341 & 4494 & 8599 & &  5255045116921270656 \\
25. &  155.7505009 -59.6109885 & 18.58 & 18.23 & 20.19 & 1.25 & A9 & 2465 & 1709 & 3656 & 2339 & 1660 & 3484 & &  5255420428334505600 \\
26. &  155.7424831 -59.6079542 & 18.51 & 18.12 & 20.21 & 1.39 & A9 & 5411 & 3653 & 7834 & 4405 & 3266 & 6013 & &  5255420428331714304 \\
27. & 155.7274145 -59.6191855 & 18.67 & 18.31 & 20.35 & 1.45 & A9 & 2716 & 1631 & 4117 & 1769 & 1475 & 3482 &  &  5255045151281158016 \\
28. &  155.7092593 -59.6284747 & 19.39 & 19.04 & 21.07 & 1.61 & A3 & 11417 & 6094 & 18663 & 6113 & 2558 & 11084 &  &  5255045116921326336 \\
29. & 155.6939437 -59.6191914 & 19.48 & 19.16 & 21.16 & 1.61 & A3 & 3787 & 2513 & 5997 & 3385 & 2021 & 7021 &  &  5255045116921334784 \\
30. & 155.7418997 -59.6083104 & 19.70 & 19.36 & 21.32 & 1.45 & A2 & 4446 & 2789 & 6513 & - & - & - & &  5255420428331714432 \\
31. &  155.7056254 -59.6148039 & 19.91 & 19.50 & 21.61 & 1.61 & A8 & 6989 & 3949 & 10840 & - & - & - &  &  5255045151281131904 \\
32. &  155.6921612 -59.6215689 & 19.98 & 19.57 & 21.74 & 1.61 & A7 & - & - & - & - & - & - &  - \\
33. &  155.7146475 -59.6129394 & 20.04 & 19.65 & 21.77 & 1.61 & A7 & 4844 & 2257 & 8489 & 5342 & 3453 & 10017 &  &  5255045151282776576 \\
34. &  155.7101136 -59.6148022 & 20.15 & 19.80 & 21.88 & 1.40 & A3 & - & - & - & - & - & - &  -\\
\hline
\multicolumn{15}{l}{$^a$ Photometry, spectral types, and reddening $E(B-V)$ were determined from 2018 MUSE data. The error on the MUSE photometry is about 0.2-0.4~mag. }\\
\multicolumn{15}{l}{\phantom{$^a$} Gaia EDR3 geometric and photogeometric distances and their uncertainties (i.e., the lower ``l'' and higher ``h'' values) are from \citet{2021AJ....161..147B}.}\\
\multicolumn{15}{l}{$^b$ Listed only for sources with parallax error $<20\%$.}\\
\multicolumn{15}{l}{$^c$ HR~Car.}\\
\multicolumn{15}{l}{$^d$ SIMBAD source GSC~08612-01828, but misidentified with star \#34 in \citet{1991A&A...246..407V}.}\\
\multicolumn{15}{l}{$^e$ Foreground star, based on the spectral energy distribution.}\\
\end{tabular}
\end{sidewaystable*}

\subsection{HR~Car as part of a moving group}
\label{massivestars_HRCar}

The MUSE data allow one to construct a color-magnitude diagram to which evolutionary tracks can be compared. Figure~\ref{figure:HRdiagram} shows such a color-magnitude diagram for the 34 MUSE sources listed in Table \ref{table:spectraltypes}, assuming an identical distance of $d=4.8$~kpc and constant interstellar reddening of $E(B-V)=0.95$~mag, with $A_I/A_V =0.60$ and $R_V=3.1$ \citep{1989ApJ...345..245C,1994ApJ...422..158O}. Two of the brightest stars (3.\ and 5.) are certainly foreground stars, as can also be inferred from their spectral energy distributions (Figure~\ref{figure:MUSEspectra}), and estimated lower reddening and much smaller Gaia distances (Table \ref{table:spectraltypes}).

Figure~\ref{figure:HRdiagram}  also shows a color-magnitude diagram using the Gaia EDR3 $G$, $G_{BP}$,  and $G_{RP}$ photometry of the MUSE sources, assuming an identical distance of $d=4.8$~kpc and a constant reddening of $E(B-V)=0.95$~mag, $R_V=3.1$, $A_G = 0.86 \times A_V$,  $A_{BP} = 1.06 \times A_V$, and $A_{RP} = 0.65 \times A_V$.\footnote{Listed on the PARSEC isochrone website \url{http://stev.oapd.inaf.it/cgi-bin/cmd}.} 
The Gaia color-magnitude diagram confirms the MUSE photometry as all sources lie in similar relative locations in both diagrams and also with respect to the overlaid theoretical single star stellar evolution isochrones from the PAdova and TRieste Stellar Evolution Code (PARSEC; \citealt{2012MNRAS.427..127B}). In addition to the MUSE sources, the positions of all Gaia sources within a radius of 3\arcmin\ ($\sim4$~pc at a distance of 4.8~kpc) around HR~Car are shown. 

While there appears to be a tight relation along a main sequence for several of the hot MUSE sources, not all stars are at the same distance and/or same reddening. For example, the faintest sources are too faint and too red for their spectral type and are thus likely background stars. 
The geometric distances and their uncertainties (\citealt{2021AJ....161..147B}; see Table~\ref{table:spectraltypes}) indicate that nine of the MUSE sources could be foreground stars with higher limits on their distances of $d_\mathrm{geo\_h} < 4448$~pc and two sources could be background stars with lower limits on their distances of $d_\mathrm{geo\_l} > 5098$~pc. At face value, the B4 and B5-type stars lie at the same distance as HR~Car, while the B0-type star is located slightly in the foreground. 
For a distance of $d=4.8$~kpc, the absolute magnitudes M$_\mathrm{V}$ for the stars with spectral type B0, B4, and B5 are $0.6$~mag, $1.3$~mag, and $0.5$~mag fainter compared to the (general) work on absolute magnitudes of OB stars based on Hipparcos parallaxes by \citet{2006MNRAS.371..185W}. However, taking into account the errors on our photometry and the distance, the error on our M$_\mathrm{V}$ estimates is on the order of 0.7~mag  and \citet{2006MNRAS.371..185W} reported errors larger than 1~mag. Deriving M$_\mathrm{V}$ using the reddening and Gaia distances from Table~\ref{table:spectraltypes} does not result in a better match with \citet{2006MNRAS.371..185W}. 
Overall, the color-magnitude diagrams and Gaia distances indicate that several of the MUSE objects, especially the B4 and B5-type stars, may be spatially associated with HR~Car.

HR~Car's bolometric luminosity and location near the 10~Myr single-star isochrone correspond to an evolutionary path of a single star with $M_{init} = 25-30~M_{\odot}$ and is consistent with the stellar mass of $M \sim 21~M_\odot$ determined by Boffin et al.\ (in prep.) by means of resolved interferometry and binary reflex motion. 
In a single-star evolution scenario, the primary has lost mass as a blue supergiant, potentially as a red supergiant, and now as an
    LBV. The circumstellar nebula contains several solar masses of material, some  mass may have escaped the system, and a fraction may have been deposited on the secondary.
HR~Car is a binary with a mass ratio of about 0.5, the LBV being the primary component (\citealt{2016A&A...593A..90B}; Boffin et al., in prep.). In an extreme case of a companion with equal $V$-band brightness, the LBV in HR~Car would be only 0.75~mag fainter and the conclusion from its location in the HR diagram would not significantly change. However, the large radius determined for the less massive secondary is in conflict with this simple single-star evolution model (\citealt{2016A&A...593A..90B}, Boffin et al., in prep.).

The uncertainties in the Gaia distances still need to be reduced to derive precise ages and masses of the stars and to determine if HR~Car's location is unexpected compared to other potential members of a moving group. For example, a better distance estimate is required to settle the question of a spatial association of the B0-type star with HR~Car.  
For a discussion on a potential past merger in a triple system or past mass transfer events, leading to a rejuvenation of HR~Car, the companion's nature needs to be identified. The large radius for the companion star indicates a red supergiant nature. However, no spectral features of a red supergiant were observed yet in near-infrared spectra, which may be due to insufficient S/N in the spectra as the luminosity ratio in $H$-band is $6-9$ (\citealt{2016A&A...593A..90B}, Boffin et al., in prep.). Radio data can also not rule out that the nebula is ionized by a hot companion star and a B0 V companion without much dust is also consistent with the data \citep{2000ApJ...539..851W}. 

To further investigate the environment of HR~Car, we analyzed the stellar content of massive stars in its environment retrieved from Simbad and Gaia in a radius of $1.2\degree$ ($100$~pc at 4.8~kpc) around the star. Figure~\ref{figure:HRclusterSimbad} provides an overview of the results. The figure shows the spatial distribution of O- and B-type stars around HR~Car with arrows indicating their movements projected on sky over the past $10^5$~yr based on Gaia proper motions. No filtering was performed based on distances. HR~Car moves in the general direction with most of the other hot stars in this region, suggestive of a co-moving group. 
While the region is rich with late O-type and B-type stars, there is only one early O-type star (i.e., earlier than O6). As stated earlier, this is not surprising given the relatively low stellar masses in the HR~Car system and its evolved nature. However, many of the OB-type star classifications originate from the VST Photometric H$\alpha$ Survey (VPHAS+; \citealt{2017MNRAS.465.1807M}) and still lack precise spectral typing. The cumulative distributions of the projected separation to O-type stars and B-type stars are similar. 

Figure~\ref{figure:HRclusterSimbad} also shows a vector point diagram, indicating with circles two standard deviations of the mean of the proper motions of early O-type, all O-type, OB-type, early B-type (i.e., earlier than B6), and B-type stars. To estimate conservatively the dispersion of the proper motion of the stars, we discarded all stars with proper motion measurements with errors $> 1$~mas~yr$^{-1}$, which removed double or multiple stars and high-proper motion stars from the sample.
HR~Car resides within two standard deviations of the mean of the proper motions of B- and O-type stars and visually appears to be part of a group, not surprising for stars at the same distance in a spiral arm. This fact, however, suggests that HR~Car is not isolated and has not received a large kick in the past from a companion SN or from a merger. Extrapolation of the proper motions over a lifetime of 10~Myr does not show convergence to a potential birth cluster, but also not divergence.

\section{The circumstellar nebula}
\label{circumstellarnebula}

Previous morphological and kinematic studies have found a bipolar structure, favoring an expanding-lobe model \citep{1997A&A...320..568W,1997A&A...321L..21V,1997ApJ...486..338N}. HR~Car's nebula is reminiscent of the Homunculus nebula of $\eta$~Car, albeit of larger size and much older ($4000-9000$~yr compared to 200~yr; \citealt{2012ASSL..384..171W,1997A&A...320..568W}). 
On the other hand, \citet{2017MNRAS.465.4147B} proposed a jet-precession model based on mid-infrared observations and the argument that the lobes do not appear complete, but S-shaped.

The MUSE velocity channel maps revealed a bipolar structure for HR~Car's large circumstellar nebula (Figure \ref{figure:channelmaps}). One lobe is in the southeast with velocities up to $-153$~km~s$^{-1}$ and one lobe is in the northwest with velocities up to $189$~km~s$^{-1}$. 
The nebula is quite fragmented and the brightness distribution is partly suggestive of an S-shaped nebula, which could have resulted from precession. However, the velocities in this comparatively old nebula show a bipolar morphology, confirming the expanding-lobe model. The observed brightness distribution may thus imply unequal mass and density distribution in the nebular shell and the surrounding medium, as well as dust clumps in the circumstellar nebula.
The companion may have played a  role in the origin and/or shaping of the bipolar nebula, as it is aligned perpendicular to the orbit. On the other hand, HR~Car rotates at $\sim0.9$ of its critical velocity \citep{2009ApJ...705L..25G} and fast-rotating massive stars can have latitude-dependent, polar-enhanced winds \citep{1996ApJ...472L.115O,2001A&A...372L...9M}. Thus, rotation could also have played a role in the star's instability and in the formation of its bipolar nebula. Possible explanations for its fast rotation could be the accretion of angular momentum through mass transfer via stable wind Roche-lobe overflow \citep{2007ASPC..372..397M} from a red supergiant companion, but also a merger in a triple system would lead to a fast rotating star.

The similarities to $\eta$~Car's Homunculus nebula are striking, which suggests similar ejection and/or shaping mechanisms. Both objects have a bipolar nebula that is aligned with the orbit of the companion star. In addition, there is also a large structure around the pinched ``waist'' of the bipolar lobes of HR~Car's nebula at low velocities that could be interpreted as an equatorial  ``skirt''-like feature (Figure~\ref{figure:channelmaps}), similar to the one observed in $\eta$~Car \citep{1999A&A...344..211Z}. Competing models for the eruptive mass loss that could lead to such nebula include super-Eddington winds (e.g., \citealt{2017MNRAS.472.3749O}),  accretion onto a companion star (e.g., \citealt{2010ApJ...723..602K}), pulsational pair-instability events (e.g., \citealt{2017ApJ...836..244W}), and the merger of two massive stars (e.g., \citealt{2016MNRAS.456.3401P,2021MNRAS.503.4276H}).
However, HR~Car is a factor of at least five times less massive in terms of both stellar and nebular mass and the orbital parameters are substantially different. The $\eta$~Car system has a highly eccentric orbit ($e\sim 0.85$) and has a complicated colliding wind system that collapses at periastron when the stars have a separation of about 1~au (e.g., \citealt{2014ApJ...784..125H}). HR~Car's orbit is nearly circular with a minimum separation between the stars on the order of 10~au (Boffin et al., in prep.), which makes stellar interactions unlikely during all but a red supergiant phase of either component.

\section{A fast bipolar outflow and ``bullets''}
\label{bipolaroutflow}

\begin{figure*}
\centering
\resizebox{1\hsize}{!}
{\includegraphics[width=0.45\textwidth]{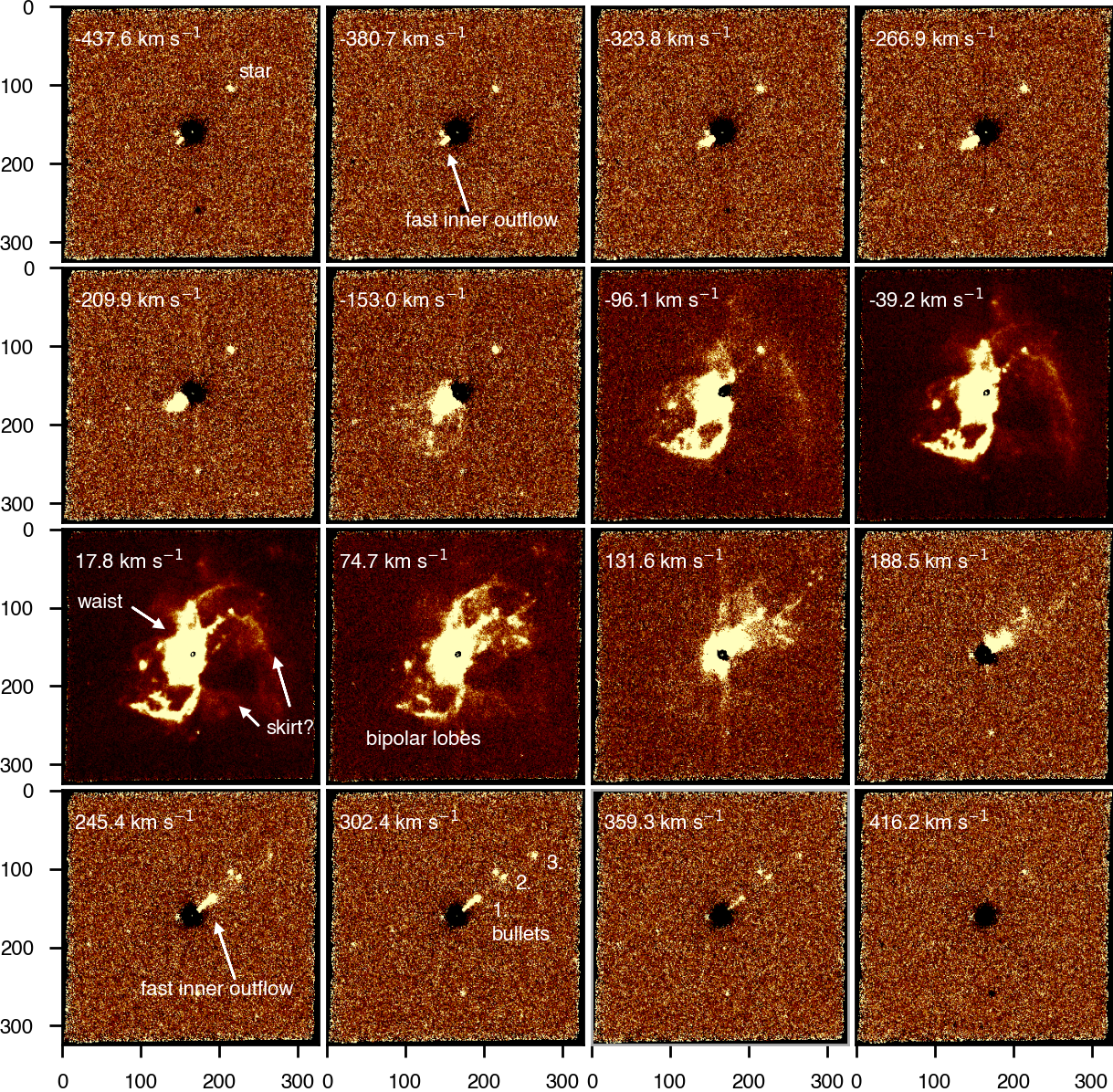}}
     \caption{MUSE channel maps showing the NII~$\lambda$6583 emission around HR~Car, accounting for a systemic velocity of $-19$~km~s$^{-1}$. The stellar continuum is subtracted. North is up, East is left, and each panel shows a region of $1\arcmin \times 1\arcmin$ on sky (the units on the axis are in pixels with $0.2\arcsec$ per pixel). In addition to a large bipolar nebula, a ``skirt''-like feature is visible at low velocities. A fast outflow extends up to 5.8\arcsec\ at a position angle on sky of $129\degree$, aligned with the geometry of the large bipolar nebula. At velocities of about 300~km~s$^{-1}$, three bullets in the northwest follow the linear trajectory of the inner outflow (labeled 1., 2., 3.; near bullet 2.\ is a star).}
     \label{figure:channelmaps}
\end{figure*}

The deep MUSE IFU observations allow one to characterize the circumstellar material around HR~Car in more detail than before  \citep{1997A&A...320..568W,1997ApJ...486..338N}.
We detect an inner fast outflow in the channel maps with velocities of up to $-440$~km~s$^{-1}$ in the southeast and up to $360$~km~s$^{-1}$  in the northwest, extending $5.8\arcsec$ from HR~Car (Figure \ref{figure:channelmaps}). The outflow has a position angle on sky of $129\degree$, similar to the slower and much larger bipolar nebula. The size of the outflow of $\sim 12\arcsec$ corresponds to a deprojected size of $0.3$~pc at a distance of $d=4.8$~kpc (see below). This inner outflow is thus similar in size and velocity as the Homunculus nebula of $\eta$~Car. 
In the northwest, there are three spatially unresolved (i.e., $<0.02$~pc at a distance of $d=4.8$~kpc) knots or ``bullets'' on a linear trajectory at  $5.8\arcsec$, $14.6\arcsec$, and $24.5\arcsec$ from the star, with the same position angle on sky as the inner outflow (labeled ``1.'', ``2.'', `3.'' in Figure \ref{figure:channelmaps}; the one labeled ``1.'' is right at the end of the fast outflow). From their nebular emission spectra we derived radial velocities of about $300$~km~s$^{-1}$, similar to the velocity of the fast outflow. Unfortunately, the S/N of the spectra is too low to further analyze these knots.

The binary orbit has a position angle of $40\degree$ (Boffin et al., in prep.). Thus, the fast inner bipolar outflow and the bullets are perpendicular to the binary orbit, suggestive of recurrent angular momentum driven accretion events. A potential red supergiant companion could account for discrete mass-loss events. However, the colliding wind shock formed by the two strong stellar winds may shield the primary from accretion.
Boffin et al.\ (in prep.) obtained an orbital inclination of $i = 119\degree$, which translates to an inclination of $29\degree$ for a potential perpendicular outflow. 
Assuming the bullets are perpendicular to the orbital plane, their velocities and physical distance are calculated as $v = v_{\mathrm{rad}}/ \mathrm{sin}\,(i-90\degree)$ and $d_{\mathrm{phys}} = 2d\,\mathrm{tan}\, (\delta/2) / \mathrm{cos}\,( i-90\degree)$, with $\delta$ the angular distance on sky (i.e., $5.8\arcsec$, $14.6\arcsec$, and $24.5\arcsec$) and $d= 4.8$~kpc.
This results in deprojected velocities of $v = 620$~km~s$^{-1}$ and physical distances of 0.15~pc, 0.38~pc, and 0.65~pc. The kinematic ages, $\tau =d_\mathrm{phys} /v$, for the three bullets are about 240~yr, 600~yr, and 1030~yr. That is, they appear to have been ejected at  intervals of about 400 years, which could be a tracer of continued past accretion in the system. Moreover, because the bullets fall on a straight line, there is no underlying precession. This is also a relevant fact for the origin of the larger bipolar nebula, discussed in Section \ref{circumstellarnebula}, as \citet{2017MNRAS.465.4147B} proposed a jet-precession model for its formation. Interestingly, these knots are only seen on one side of the star.

The bullets may be similar to the five so-called strings observed around $\eta$~Car, which are long, highly collimated linear features that appear to trace back to the central star \citep{1999A&A...349..467W}. The physical nature of $\eta$~Car's strings is not understood, but points toward the Great Eruption that created the Homunculus nebula. 
However, for $\eta$~Car the strings follow a Hubble law with velocities increasing toward larger distances from the star, while for HR~Car all three bullets have the same velocity, independent of distance. 
Similarities can also be drawn to a collimated outflow with projected velocities of up to $\pm450$~km~s$^{-1}$ centered on the B[e] star MWC~137 \citep{2016A&A...585A..81M}.

\section{Discussion}
\label{environment_LBVs}

Our work on the individual object HR~Car cannot resolve the debate if LBVs are the descendants of early O-type stars evolving as single stars, rejuvenated mass gainers, or the result of mergers. 
However, we provide an assessment of HR~Car's spatial relationship to O- and B-type stars and its potential membership in a moving group that can be translated to other LBVs. 

From the 34 hot stars in the close vicinity of
HR~Car  ($r < 2\arcmin \sim 2.8$~pc), identified in the MUSE data, not one object had a
spectral type classification cataloged in Simbad at the time of
this study.  Our MUSE data underline the
sparse knowledge of stellar classifications even around well-studied stars like HR~Car. This indicates a general problem in the assessment of
the degree of isolation of LBVs. Similar deep population studies need to be
conducted before a relatively unbiased measure is obtained on the
degree of their isolation.

Existing catalogs of O-type and early B-type stars are confined to objects brighter than $V\sim12$~mag and complete only to $\sim2$~kpc. A case in point is the recent work by \citet{2021A&A...648A..34P}, who performed an infrared spectroscopic study of the population of the young stellar cluster TR~16-SE in the Carina Nebula. The authors increased the number of spectroscopically identified early-type O5--B2 stars in Tr 16-SE from two to nine. Infrared spectroscopy was required to reveal the hidden population of early-type stars in this obscured young massive cluster. 
Data from the VST Photometric H$\alpha$ Survey (VPHAS+) of the Carina Arm over the Galactic longitude range  $282\degree \leq l \leq 293\degree$ revealed 5915 O-type to B2-type star candidates and 5170 stars with a later B spectral type and also a so far unnoticed association of 108 O-type to B2-type stars \citep{2017MNRAS.465.1807M}.
Census incompleteness can therefore be a significant factor in assessing LBV isolation.  

An additional complication in assessing the degree of isolation
of LBVs relates to the mode of star formation itself and the
stochasticity of initial mass function sampling. A relation between most
massive star and cluster mass would indicate that stellar clusters are populated from the stellar initial mass function by random sampling over the mass range. However, there appears to be a dependence of the stellar inventory of a star cluster on its mass, which may be due to the interplay between stellar feedback and the binding energy of the cluster-forming molecular cloud core \citep{2010MNRAS.401..275W}. More recent work brings tension to how closely related stellar clusters and massive-star formation are.  \citet{2020A&A...643A.138M} investigated 16
Galactic O-type stellar groups each containing O2--O3.5 stars. The
authors concluded that very massive (binary) stars can
form in relatively low-mass clusters or even in near-isolation.  While
some stars are born in well-defined bound clusters, others are born in
unbound associations. 

Nonetheless, as the isolated formation of massive stars would be a rare occurrence, the majority of high-mass LBVs should be associated with a high-mass stellar group if they have not received large kicks from a companion SN or a merger event. On the other hand, lower-mass LBV systems such as HR~Car are not expected to be in OB associations as most O-type stars would have already ended their lives. For the lower-mass LBVs, membership within a moving group can be established by means of a shared proper motion vector (e.g., Figure \ref{figure:HRclusterSimbad}). This method can also be applied to higher-mass LBVs and complements and can contradict group identification by mere spatial association.

\section{Conclusions}
\label{conclusion}

The suggestion that LBVs practice ``social distancing'', that is are single stars located away from young OB association argues in favor of a binary origin of the LBV phenomenon and against single-star evolution \citep{2015MNRAS.447..598S}. In this paper, we have turned the question if LBVs are accompanied by O-type stars into the question if LBVs are alone. 
The masses of the two binary components in the Galactic LBV HR~Car and the mass of the surrounding nebula indicate that the initial stellar masses in this system barely exceeded the O-star threshold. One would thus not expect any early O-type stars or many late O-type stars associated with HR~Car. 
Instead, we have shown that HR~Car is not isolated, but may reside within a moving group  (in a spiral arm) of about $10-30$~Myr, with a main sequence populated up to B0 or B4. Proper motion analysis confirms that HR~Car has not received an extraordinary large space velocity kick owing to a companion SN or merger event. 

HR~Car is an interferometrically resolved binary, which experienced recurrent outflow events perpendicular to its binary orbit (bipolar nebula, fast outflow, ``bullets''). Especially, the ``bullets'' indicate recurrent past activity related to either rapid rotation or binarity. The bolometric luminosity of HR~Car and its location in the color-magnitude diagram is in general agreement with a single-star evolution of a star with an initial mass of $M_\mathrm{init} = 25-30~M_\odot$, which has lost mass during a blue supergiant, potentially a red supergiant, and currently a LBV phase. However, the unsolved question of the nature of the companion star is cause of caution to adopt this simple model. The large radius derived from interferometric observations suggests a red supergiant nature.
In an extreme case where HR~Car would have undergone rejuvenation by mass transfer from a red supergiant companion (potentially via wind Roche-lobe overflow), the secondary star must have been stripped of its outer layers quite dramatically and lost much of its mass. If the companion's mass is now 50\% of that of the primary, the secondary may have lost 50\% of its original mass or more. However, such a scenario is not easy to reconcile with the current orbital parameters and detailed binary evolution modeling is needed. Based on our findings for this individual LBV, we cannot rule out that mass stripping and rejuvenation may still be viable explanations for generating the LBV phenomenon, as may be  companion SNe or merger events in a triple system without imparting a large space velocity kick.

\begin{acknowledgements} 
This research has made use of the SIMBAD database,
operated at CDS, Strasbourg, France \citep{2000A&AS..143....9W}.
This work has made use of data from the European Space Agency (ESA) mission
{\it Gaia} (\url{https://www.cosmos.esa.int/gaia}), processed by the {\it Gaia}
Data Processing and Analysis Consortium (DPAC,
\url{https://www.cosmos.esa.int/web/gaia/dpac/consortium}). Funding for the DPAC
has been provided by national institutions, in particular the institutions
participating in the {\it Gaia} Multilateral Agreement.
The national facility capability for SkyMapper has been funded through ARC LIEF grant LE130100104 from the Australian Research Council, awarded to the University of Sydney, the Australian National University, Swinburne University of Technology, the University of Queensland, the University of Western Australia, the University of Melbourne, Curtin University of Technology, Monash University and the Australian Astronomical Observatory. SkyMapper is owned and operated by The Australian National University's Research School of Astronomy and Astrophysics. The survey data were processed and provided by the SkyMapper Team at ANU. The SkyMapper node of the All-Sky Virtual Observatory (ASVO) is hosted at the National Computational Infrastructure (NCI). Development and support the SkyMapper node of the ASVO has been funded in part by Astronomy Australia Limited (AAL) and the Australian Government through the Commonwealth's Education Investment Fund (EIF) and National Collaborative Research Infrastructure Strategy (NCRIS), particularly the National eResearch Collaboration Tools and Resources (NeCTAR) and the Australian National Data Service Projects (ANDS).
This research has made use of NASA's Astrophysics Data System Bibliographic Services.
SJ acknowledges support from the FWO PhD fellowship under project 11E1721N.
JB acknowledges support from the FWO\_Odysseus program under project G0F8H6N.

\end{acknowledgements}

\bibliographystyle{aa}

\end{document}